\documentstyle[11pt,aaspp4]{article}

\slugcomment{To Appear in The Astronomical Journal}

\lefthead{Schneider et al.}
\righthead{High-Redshift Quasars}

\begin{document}

\title{High-Redshift Quasars Found in Sloan Digital Sky Survey Commissioning
 Data~V.  Hobby-Eberly Telescope Observations
\footnote{Based on observations obtained with the Sloan Digital
 Sky Survey, which is owned and operated by the Astrophysical Research
 Consortium.}$^,$\footnote{Based on observations obtained with the Hobby-Eberly
 Telescope, which is a joint project of the University of Texas at Austin,
 the Pennsylvania State University, Stanford University,
 Ludwig-Maximillians-Universit\"at M\"unchen, and Georg-August-Universit\"at
 G\"ottingen.}
}
\author{
Donald~P.~Schneider\altaffilmark{\ref{PennState}},
Xiaohui~Fan\altaffilmark{\ref{Princeton}}$^,$\altaffilmark{\ref{IAS}},
Michael~A.~Strauss\altaffilmark{\ref{Princeton}},
James~E.~Gunn\altaffilmark{\ref{Princeton}},
Gordon~T.~Richards\altaffilmark{\ref{PennState}},
Gary~J.~Hill\altaffilmark{\ref{Texas}},
Phillip~J.~MacQueen\altaffilmark{\ref{Texas}},
Lawrence~W.~Ramsey\altaffilmark{\ref{PennState}},
Mark~T.~Adams\altaffilmark{\ref{Texas}},
John~A.~Booth\altaffilmark{\ref{Texas}},
Grant~M.~Hill\altaffilmark{\ref{Texas}},
G.R.~Knapp\altaffilmark{\ref{Princeton}},
Robert~H.~Lupton\altaffilmark{\ref{Princeton}},
David~H.~Saxe\altaffilmark{\ref{IAS}},
Matthew~Shetrone\altaffilmark{\ref{Texas}},
Joseph~R.~Tufts\altaffilmark{\ref{Texas}},
Daniel~E.~Vanden~Berk\altaffilmark{\ref{FNAL}},
Marsha~J.~Wolf\altaffilmark{\ref{Texas}},
Donald~G.~York\altaffilmark{\ref{Chicago}},
John~E.~Anderson,~Jr.\altaffilmark{\ref{FNAL}},
Scott~F.~Anderson\altaffilmark{\ref{UW}},
Neta~A.~Bahcall\altaffilmark{\ref{Princeton}},
J.~Brinkmann\altaffilmark{\ref{APO}},
Robert~Brunner\altaffilmark{\ref{Caltech}},
Istvan~Csab\'ai\altaffilmark{\ref{JHU}}$^,$\altaffilmark{\ref{Hungary}},
Masataka~Fukugita\altaffilmark{\ref{CosJapan}}$^,$\altaffilmark{\ref{IAS}},
G.S.~Hennessy\altaffilmark{\ref{USNODC}},
\v Zeljko~Ivezi\'c\altaffilmark{\ref{Princeton}},
Donald~Q.~Lamb\altaffilmark{\ref{Chicago}},
Jeffrey~A.~Munn\altaffilmark{\ref{USNOAZ}},
and
Aniruddha~R.~Thakar\altaffilmark{\ref{JHU}}
}

email addresses: dps@astro.psu.edu, fan@sns.ias.edu,
strauss@astro.princeton.edu

\newcounter{address}
\setcounter{address}{3}
\altaffiltext{\theaddress}{Department of Astronomy and Astrophysics, The
   Pennsylvania State University, University Park, PA 16802.
\label{PennState}}
\addtocounter{address}{1}
\altaffiltext{\theaddress}{Princeton University Observatory, Princeton,
   NJ 08544.
\label{Princeton}}
\addtocounter{address}{1}
\altaffiltext{\theaddress}{The Institute for Advanced Study, Princeton,
   NJ 08540.
\label{IAS}}
\addtocounter{address}{1}
\altaffiltext{\theaddress}{Department of Astronomy, McDonald Observatory,
   University of Texas, Austin, TX~78712.
\label{Texas}}
\addtocounter{address}{1}
\altaffiltext{\theaddress}{Astronomy and Astrophysics Center, University of
   Chicago, 5640 South Ellis Avenue, Chicago, IL 60637.
\label{Chicago}}
\addtocounter{address}{1}
\altaffiltext{\theaddress}{Fermi National Accelerator Laboratory, P.O. Box 500,
   Batavia, IL 60510.
\label{FNAL}}
\addtocounter{address}{1}
\altaffiltext{\theaddress}{University of Washington, Department of
   Astronomy, Box 351580, Seattle, WA 98195.
\label{UW}}
\addtocounter{address}{1}
\altaffiltext{\theaddress}{Apache Point Observatory, P.O. Box 59,
   Sunspot, NM 88349-0059.
\label{APO}}
\addtocounter{address}{1}
\altaffiltext{\theaddress}{Astronomy Department, California Institute of
   Technology, Pasadena, CA 91125.
\label{Caltech}}
\addtocounter{address}{1}
\altaffiltext{\theaddress}{Department of Physics and Astronomy,
   Johns Hopkins University, 3701 University Drive, Baltimore, MD 21218.
\label{JHU}}
\addtocounter{address}{1}
\altaffiltext{\theaddress}{Department of Physics of Complex Systems,
   E\"otv\"os University, P\'azm\'ay P\'eter \hbox{s\'et\'any 1/A,}
   H-1117, Budapest, Hungary.
\label{Hungary}}
\addtocounter{address}{1}
\altaffiltext{\theaddress}{Institute for Cosmic Ray Research, University
   of Tokyo, Midori, Tanashi, Tokyo 188-8588, Japan
\label{CosJapan}}
\addtocounter{address}{1}
\altaffiltext{\theaddress}{US Naval Observatory, 3450 Massachusetts Avenue NW,
   Washington, DC 20392-5420.
\label{USNODC}}
\addtocounter{address}{1}
\altaffiltext{\theaddress}{US Naval Observatory, Flagstaff Station,
   P.O. Box 1149, Flagstaff, AZ 86002-1149.
\label{USNOAZ}}

\vbox{
\begin{abstract}
We report the discovery of 27 quasars with redshifts between~3.58 and~4.49.
The objects were identified as high-redshift candidates
based on their colors in Sloan Digital Sky Survey commissioning data.
The redshifts were confirmed with low resolution spectra obtained at the
Hobby-Eberly Telescope.
The quasars' $i^*$ magnitudes range from~18.55 to~20.97.
Nearly~60\%
of the quasar candidates observed are confirmed spectroscopically as quasars.
Two of the objects are Broad Absorption Line quasars, and several other
quasars appear to have narrow associated absorption features.  

\end{abstract}
}



%

\section{Introduction}

The Sloan Digital Sky Survey (SDSS; \hbox{York et al.~2000)} has proven to
be a rich source of high-redshift \hbox{($z \ge 3.5$)} quasars; to date,~86
high-redshift
quasars have been published by the SDSS, including four that have
redshifts larger than~4.95 (\hbox{Fan et al. 1999,} Paper~I; 2000a,
Paper~II; 2000b; 2001a, Paper~III;
\hbox{Schneider et al. 2000a,b;} \hbox{Zheng et al. 2000}).
In this paper we report the discovery of 27 quasars
at \hbox{$3.5 < z < 4.5$} taken from SDSS commissioning data along the
Celestial Equator.  High-redshift quasar candidates were identified using
a multicolor selection technique ($e.g.,$ Warren, Hewett, and Osmer~1994).
Low-resolution spectra of the candidates were obtained with
the Hobby-Eberly Telescope (HET; Ramsey~et~al.~1998, Hill~2000)
during its first year of science operations.
The data were acquired as part of our investigation of the high-redshift
quasar
luminosity function using SDSS data.  Eight of the objects discussed in this
paper (many of the brighter ones located in the Southern Galactic sky)
form part of the
complete sample of high-redshift quasars discussed in Paper~III and by
\hbox{Fan et al. (2001b, Paper~IV)}.  The remaining quasars do not form
a complete sample,
but are presented at this time so the community can have rapid access to the
sources, several of which have interesting absorption features.

The SDSS and HET observations are described in \S 2, the properties of
the quasars presented in \S 3, and a brief discussion appears in \S 4.
Throughout this paper we will adopt the cosmological model with
\hbox{$H_0$ = 50 km s$^{-1}$ Mpc$^{-1}$,} \hbox{$\Omega_0$ = 1.0,}
\hbox{and $\Lambda$ = 0.0.}
\section{Observations}

\subsection{Sloan Digital Sky Survey} \label{survtech}

The Sloan Digital Sky Survey
uses a CCD camera \hbox{(Gunn et al. 1998)} on a
dedicated 2.5-m telescope \hbox{(Siegmund et al. 2001)}
at Apache Point Observatory,
New Mexico, to obtain images in five broad optical bands over
10,000~deg$^2$ of the high Galactic latitude sky centered approximately
on the North Galactic Pole.  The five filters (designated $u'$, $g'$,
$r'$, $i'$, and~$z'$) cover the entire wavelength range of the CCD
response \hbox{(Fukugita et al. 1996; Paper~III)}.
Photometric calibration is provided by simultaneous
observations with a 20-inch telescope at the same site.  The
survey data processing software measures the properties of each detected object
in the imaging data, and determines and applies astrometric and photometric
calibrations (\hbox{Pier et al. 2001}; \hbox{Lupton et al. 2001}).

The high photometric accuracy of the SDSS images and the information
provided by the $z'$ filter (central wavelength of 8873~\AA , see Paper~III)
make the
SDSS data an excellent source for identification of high-redshift
quasar candidates.  The SDSS high-redshift quasar selection efficiency
(number of quasars divided by the number of quasar candidates) is
approximately \hbox{60--70\%}
\hbox{($e.g.,$ Papers I, II, and III)}
for sources with \hbox{$i^* < 20.0$,} which is a much
higher value than that achieved in previous investigations in this field.

We have started a survey of faint, high-redshift quasars using the SDSS
imaging data.  This investigation uses a multicolor selection technique
to identify suitable candidates; the locations of \hbox{$z > 3.5$} quasars
in the multidimensional SDSS color space are well separated
from the stellar locus
(see Fan~1999; Papers~I and~II).  A detailed discussion of
the selection criteria can be found in Paper~III and references
therein.  For the quasars in this paper, all having redshifts less than~4.5,
the \hbox{$(g^*-r^*, r^*-i^*)$} diagram is the key to separating the quasars
from the stellar locus.
In this study the candidates must have an~$i^*$ magnitude brighter
than~21.0; this is one
magnitude fainter than the limit imposed in Papers~I and~II, and that of
the complete sample in Papers~III and~IV.  We are currently assembling
a sample of quasars down to a magnitude limit \hbox{$i^* = 21$} to investigate
the \hbox{$z > 3.5$} quasar luminosity function luminosities considerably
fainter than that reached in Paper~IV.

The quasars presented here were identified from four different SDSS scans
along the Celestial Equator (observation dates in parentheses):
\hbox{94 (1998 September 19),} \hbox{125 (1998 September 25),}
\hbox{745 (1999 March 20),}
\hbox{and 756 (1999 March 22).}  All of the data were taken under photometric
conditions, and the typical seeing was~1.5$''$.  The total area covered is
approximately~500~square~degrees.

Notes on SDSS nomenclature:  The source name format is
\hbox{SDSSp Jhhmmss.ss+ddmmss.s}, where the coordinate
equinox is~J2000, and the ``p" refers to the preliminary nature of the
astrometry.  The reported magnitudes
are based on a preliminary photometric calibration; to indicate this,
the filters have an asterisk instead of a prime superscript.  The
estimated astrometric accuracies in each coordinate are~0.15$''$ and the
calibration of the photometric measurements is accurate to~0.04~magnitudes
in the~$g'$, $r'$, and~$i'$ filters
and~0.06~magnitudes in the $u'$ and~$z'$ bands.
Throughout the text, object names will be abbreviated \hbox{as SDSShhmm+ddmm.}

\subsection{Spectroscopy of Quasar Candidates} \label{survfields}

Spectra of~50 SDSS high-redshift quasar candidates were obtained
with the HET's Marcario
Low Resolution Spectrograph (LRS; Hill~et~al.~1998a,b;
Cobos~Duenas~et~al.~1998; Schneider~et~al.~2000b) between October~1999
and June~2000.
The LRS is mounted in the Prime Focus Instrument Package, which
rides on the HET tracker.
The dispersive
element was \hbox{a 300 line mm$^{-1}$} grism blazed \hbox{at 5500 \AA .}
An OG515 blocking filter was installed to permit calibration of the spectra
beyond~8000~\AA .
The detector is a thinned, antireflection-coated
3072~$\times$~1024 Ford Aerospace CCD, and was
\hbox{binned $ 2 \times 2$} during readout; this produced
an image scale of~0.50$''$~pixel$^{-1}$ and a dispersion
\hbox{of $\approx$ 4.5 \AA\ pixel$^{-1}$}.  The spectra covered the
range from 5100--10,200~\AA\ at a resolution of approximately~20~\AA.

The wavelength calibration was provided by Ne, Cd, and~Ar comparison lamps; a
cubic fit to the lines produced an rms error of less than 1~\AA .
The relative flux calibration and atmospheric absorption band corrections
were performed by
observations of spectrophotometric standards, usually the primary
spectrophotometric standards of Oke \& Gunn~(1983). 
The objects were observed under a wide range of conditions; the FWHM of the
spectra ranged from slightly under~2$''$ to over~4$''$.  The exposure times
varied from 566~s to 1800~s, with a median of 800~s.
Absolute spectrophotometric calibration was performed
by scaling each spectrum so that the~$r^*$
and~$i^*$ magnitudes synthesized from the spectra matched the SDSS photometric
measurements; this scaling used the new SDSS response curves of Paper~III.

Of the 50 objects observed, the spectra of two had essentially no signal
(weather related), 13 had featureless spectra (usually low
S/N, but of sufficient quality to rule out a typical quasar
Lyman~$\alpha$ emission line),
one was a faint carbon star (see Margon~et~al.~2001),
two were strong, narrow emission line objects at redshifts of a few
tenths, three were E-type galaxies at redshifts between~0.4 and~0.7,
three were E+A galaxies at redshifts of~$\approx$~0.4 (in these
objects, the Balmer jump produces SDSS colors that mimic
the Lyman~$\alpha$ emission line/forest boundary),
and~28 were quasars with redshifts between~3.58 and~4.49.
One of the quasars, the brighter member of a $z$~=~4.25 pair, was
described by Schneider~et~al.~2000a.  This efficiency (28 of~48, or~58\%)
is comparable to previous SDSS studies.

Finding charts for the~27 new quasars are given in Figure~1; each panel is
the SDSS~$i'$ image of the field. (The finding chart for the $z$~=~4.25
quasar mentioned in the previous paragraph can
be found in Schneider~et~al.~2000a.)
The flux and wavelength calibrated
spectra of the objects are presented in Figure~2, where the data have
been rebinned \hbox{at 8 \AA\ pixel$^{-1}$.}
We do not show the data
beyond~9000~\AA\ as the signal-to-noise ratio in this region is low
for the vast majority of the objects.  Two of the spectra (SDSS0239$-$0021
and SDSS0839+0037) display prominent BAL features, and the sample contains
a wide variety of Lyman~$\alpha$ emission line profiles.

\section{Properties of the Quasars}

The five-band SDSS photometry for the quasars is presented in Table~1.
The far right column contains the SDSS run number of the data set from which
the object was selected as a high-redshift quasar candidate.

The redshifts were measured from the centers of one to four
emission lines (\hbox{O I + Si II,} \hbox{Si IV + O IV],}
C~IV, and~C~III]); the \hbox{Lyman~$\alpha$ + N V} emission
line was not used, as it is frequently strongly affected by absorption from the
Lyman~$\alpha$ forest.

Table~2 presents a number of properties of the quasars.  The spectra were
corrected for Galactic reddening using a standard reddening law and
the maps of Schlegel, Finkbeiner, \&~Davis~(1998); the $E(B-V)$ for
each line-of-sight is listed in column~3.  A power law was fit to
the continuum of the dereddened spectra between 1260~\AA\ and 1650~\AA\ in the
rest frame; the normalization ($AB_{1450}$, the $AB$ magnitude at rest
wavelength 1450~\AA ) and power law slope
\hbox{($f_{\nu} \propto \nu^{\alpha}$)} for each object is given in
columns~4 and~5 of Table~2.  The final column in Table~2 is the absolute
magnitude at 1450~\AA\ in the rest frame.  Absolute $B$ magnitudes can
be calculated from \hbox{$M_B = M_{1450} + 1.21 \alpha + 0.12$}, where
$\alpha$ is the power law slope between 1450~\AA\ and 4400~\AA\ in the
rest frame.  In our adopted cosmology,
3C~273 has \hbox{$M_{1450} \approx -26.5$}; the quasars range from~0.18
to 1.8 times as luminous as 3C~273.

As was mentioned in the introduction, eight of these quasars are members of
the complete sample presented in Paper~III; in that work
the $AB_{1450}$ values
and power law slopes were determined via different techniques
than those here (Paper~III matched the SDSS broad band photometry to
a continuum plus emission line model), and the redshifts were measured using
different software packages.
In general the agreement between the two methods is
quite good; the largest redshift discrepancy is~0.02 (the mean agreement
is better than~0.01), and the $AB_{1450}$ values are tightly correlated with
a systematic offset of~0.06~mag (the values in this paper are
brighter).  The two techniques yield power law slopes for seven objects that
are in excellent agreement (0.2~dispersion with no systematic offset), but
the measured
slopes for SDSS2256+0047 are quite discrepant \hbox{($-0.25$ vs. $-1.15$).}

The average ultraviolet power law slope in this sample is $-0.93$ with
\hbox{a 1$\sigma$} dispersion of 0.31; this compares to the values of
$-0.91$ and~0.26 from Schneider, Schmidt, \& Gunn~(1991, SSG91) and $-0.79$
and~0.34 from Paper~III.  (Note that the Paper~III value is not a simple
average but an estimate of the true population given the selection effects
of the sample.)
These measurements are all significantly steeper than
the canonical $-0.5$ median ($-0.6$~mean) value found by
Richstone \& Schmidt~(1980, RS80).  It should be noted that the maximum redshift
in the RS80 sample is~2.686 (the majority have redshifts below two),
so their measurements tend to be made
at considerably longer rest wavelengths than the region where we determine
the continuum properties .  RS80 did not find any evidence
for curvature in the continuum, however, so the observed steepening of the power
law with redshift cannot arise from a universal quasar continuum that
steepens as one moves to shorter rest wavelengths.  RS80 did
remark that there was a hint that
the power law index may be steepening with redshift (see RS80 Figure~8).

Natali~et~al.~(1998) did not find any evidence for a $\alpha$-$z$ relation
in their investigation of a sample of 62 radio-quiet quasars with
redshifts between~0.5 and~2.5.  They did note (see their Figure~3)
that the measured value of $\alpha$ did depend on the rest wavelength
used in the fit, in the sense that slopes measured when the rest wavelengths
were restricted below 3000~\AA\ were steeper than those that included
rest wavelengths in the optical.  A study of~73 radio-loud quasars
with \hbox{$0.4 < z < 2.8$} by Carballo~et~al.~(1999) found that the
ultraviolet power law slope in such objects appears to flatten as one
moves to higher redshift.

One possible source for the differing values of the slope is the restricted
wavelength range used in the continuum fits in high-redshift quasars,
an effect hinted at by Natali~et~al.~(1998).
As redshifts approach four, it becomes difficult to obtain accurate data
much beyond the C~IV line, and the rest wavelength
\hbox{range 1600--1700 \AA } plays a significant role in determining the
slope.  The composite quasar spectrum of Francis~et~al.~(1991) indicates the
presence of significant broad Fe emission in this region; this
feature is particularly prominent in the composite formed by
over 2200 quasar spectra obtained with the SDSS spectrographs
(Vanden~Berk~et~al.~2001).  If one measures the continuum slope in the
SDSS composite using the rest frame \hbox{region 1250--1700 \AA } and
treats the region redward of the C~IV line as continuum, one finds a
value \hbox{of $\alpha = -0.93$.}

This information suggests that the \hbox{$\alpha \approx -1.0$ to $-0.8$}
slopes in high redshift quasars may be due to the continuum regions used
in the fit and may not signify a change in spectral index with redshift.
Given this uncertainty of the continuum shape between the rest frame
ultraviolet and optical in the rest frame in \hbox{$z \approx 4$} quasars,
it is not clear how to
extrapolate~$M_{1450}$ to $M_B$; a change from~$-0.5$ to~$-0.9$ in~$\alpha$
induces an increase of more than~50\% in the inferred blue luminosity!

A striking aspect of this data set is the wide variety of
\hbox{Lyman~$\alpha$ + N V}
emission line profiles.  The rest equivalent widths of the feature range
from 30~\AA\ to~151~\AA , with a mean of 70~\AA , similar to that found
in previous studies ($e.g.$, SSG91, Papers~I,~II, and~III).
The line shapes range from barely resolved Lyman~$\alpha$
only ($e.g.,$ SDSS2303+0016), sharp, distinct Lyman~$\alpha$ and N~V peaks
($e.g.,$ SDSS0234$-$0014), sharp Lyman~$\alpha$ plus a broad red component
($e.g.,$ SDSS0139$-$0101), narrow absorption superposed on broad line
($e.g.,$ SDSS0339$-$0030), broad line with abrupt blue
absorption ($e.g.,$ SDSS0939+0039), and the two BALs (SDSS0239$-$0021 and
SDSS0839+0037).

The effect of absorption of the \hbox{Lyman~$\alpha$ + N V} feature can
be seen in the quantity $\Delta L \alpha$, which is the difference
between the rest frame peak of the Lyman~$\alpha$ emission
line and~1215.7~\AA\ (see SSG91).
Column~7 of Table~2
presents $\Delta L \alpha$ for the sample; only three of the~27 values
are negative ($i.e.,$ the peak of the Lyman~$\alpha$ line occurs shortward
of rest frame~1215.67~\AA ),
and the largest negative deviation is only~2.3~\AA.  Six of the
quasars have $\Delta L \alpha$ larger \hbox{than +10 \AA .}  The distribution
of $\Delta L \alpha$ is similar to that seen in~SSG91; the BALs have large
values of $\Delta L \alpha$, and the median $\Delta L \alpha$
of the 27~quasars is~+2.8~\AA .

The spectroscopic blue limit of 5100~\AA\ prohibits investigation of Lyman
limit systems in these quasars, and for most of the objects it is not
possible to perform a proper measurement of the continuum depression ($D_A$)
produced by the Lyman~$\alpha$ forest.  Three objects, (\hbox{SDSS1135+0024,}
\hbox{SDSS2256+0047,} and \hbox{SDSS2306+0108}), have strong absorption
features that have redshifts of~$\approx$~3.6 if they are damped
Lyman~$\alpha$ lines.

One of the quasars, SDSS0844+0018 ($z$~=~3.69), is radio-loud;
it has an observed 20~cm flux density of~6.05~mJy
in the FIRST (Becker, White \& Helfand~1995) survey.  The other~26
quasars are not detected at the 1~mJy level
and are therefore
radio-quiet, as are the vast majority of optically
selected \hbox{$z > 3$ quasars}
(Schneider et al.~1992; Schmidt et al.~1995b; Stern et al.~2000).
None of the quasars are found in the ROSAT Bright Survey Catalog
(Schwope et al.~2000),
which has a flux limit of \hbox{$2.4 \times 10^{-12}$ erg cm$^{-2}$ s$^{-1}$}
in the 0.5~to~2.0~keV band;
this is not surprising given that except for a few exceptional sources
($e.g.,$ blazars), high-redshift quasars have X-ray fluxes well below the
catalog limit (Kaspi, Brandt, \& Schneider~2000; Brandt~et~al.~2001).

\bigskip
\centerline{Notes on Individual Objects}

\medskip\smallskip\vbox{\noindent
{\bf SDSSp J023446.58$-$001415.9} ($z = 3.60$): The Lyman~$\alpha$,
N~V, and C~IV lines all have a very sharp core and broad wings.
}

\noindent
{\bf SDSSp J023908.98$-$002121.5} ($z = 3.73$): A BAL with a number of
quite impressive absorption troughs; the C~IV feature nearly reaches
zero even at this low spectral resolution.

\medskip\vbox{\noindent
{\bf SDSSp J031036.85+005521.7} ($z = 3.79$): The C~IV emission line is
split into two sections by a very deep, narrow absorption feature
that appears to be centered
slightly to the red of the center of the emission line.
The center of
this absorption nearly reaches zero flux density even at this
spectral resolution.
}

\noindent
{\bf SDSSp J033910.53$-$003009.2} ($z = 3.74$): Narrow absorption 
features are present
on the blue wing of the C~IV emission line and in the
\hbox{Lyman~$\alpha$ + N V} emission line; the absorptions are
probably produced by intrinsic C~IV and N~V, respectively.

\noindent
{\bf SDSSp J083929.33+003759.0} ($z = 3.73$): A BAL with several
strong BAL troughs.

\noindent
{\bf SDSSp J230323.77+001615.2} ($z = 3.70$): The spectrum shows
strong, barely resolved \hbox{Lyman~$\alpha$ + N V} and C~IV lines.

\noindent
{\bf SDSSp J230639.65+010855.2} ($z = 3.64$): This is probably another
example of a quasar with associated N~V and~C~IV absorpton.

\section{Discussion}

The results
in this paper bring the total number of published \hbox{$z > 3.5$} quasars
identified by the SDSS to~113; 23 have redshifts larger than~4.4.  (Six of
the 113 objects were previously known, but the SDSS quasar selection
algorithm identified them as high-redshift quasar candidates.  We will
include the six objects in the discussion and figures that follow.
We do not include the fainter quasar of \hbox{the $z = 4.25$} pair
serendipitously found by the SDSS [Schneider~et~al.~2000a] as this object
did not satisfy our selection criteria.)
Figure~3 shows the locations of these quasars in the
\hbox{$(g^*-r^*),(r^*-i^*)$} and \hbox{$(r^*-i^*),(i^*-z^*)$}
color-color diagrams.
The figure also displays the colors of stars with \hbox{$i^* < 20.0$} from
\hbox{25 square degrees} of SDSS imaging data and the expected
colors of quasars as a function of redshift (solid line; see Fan~1999).
Note that the quasars
have been found in a region that covers less than~10\% of the planned
SDSS area.

Figure~3 shows that the ``$g^* r^* i^*$'' diagram (left panel)
is very effective at
separating quasars with \hbox{$3.5 < z < 4.5$} from the stellar locus, and
the observed colors of these objects follow the expected relation reasonably
well.  As the redshifts approach~4.5, Lyman limit systems can enter
(and dominate) the $g^*$ measurements; this leads to a large dispersion
in \hbox{the $(g^* - r^*)$} colors in such sources, and quasars with
redshifts greater than~4.5 are frequently undetected in~$g^*$.  (Note that
there are only four quasars \hbox{with $z > 4.5$} in the left panel of
Figure~3.)
Quasars with redshifts lower than~$\approx$~4.5
are not separated from the stellar
locus in the ``$r^* i^* z^*$" diagram (right panel of Figure~3) as at this
redshift the Lyman~$\alpha$ forest is just starting to have a significant
effect on the~$r^*$ measurement.  These effects combine to lower
the selection efficiency of \hbox{$z \approx 4.5$} quasars (see Paper~III
for a detailed discussion).
Regarding this point,
\hbox{SDSS0939+0039,} the $z$~=~4.49 quasar presented in this paper, is the
first SDSS quasar found in the \hbox{4.45-4.55} redshift bin.   At redshifts
above~$\approx$~4.6 the Lyman~$\alpha$ forest begins to depress the~$r^*$
flux, and the quasar colors in the $r^* i^* z^*$ are quite distinct from
the stellar locus (right panel).

The redshift distribution of the SDSS quasars is displayed in Figure~4.
The main features are the sudden cutoff below \hbox{$z \approx 3.6$}, which
arises because of our reliance on the ``$g^*r^*i^*$" color-color plot
(lower redshift quasars are found using the ``$u^*g^*r^*$" diagram; see
Richards~et~al.~2001), the decline above $z \approx 3.7$ due to the
combination of the rapid fall of the quasar luminosity function above
$z \approx 3$ ($e.g.$, Schmidt~et~al.~1995a, Paper~IV, and references therein)
and the simple fact that objects become fainter as the redshift increases,
and the dip at \hbox{$z \approx 4.5$} discussed in the previous paragraph.

The results here and in Paper~III indicate
that the SDSS will be able to efficiently identify
\hbox{$z \approx 4$} quasars at luminosities approximately one magnitude
fainter than~3C~273.  Paper~III and this paper mark the start of our
attempt to assemble a complete sample of
high-redshift quasars to study the quasar luminosity function at
considerably lower luminosities than that in Paper~IV, as well as
investigate clustering of high-redshift quasars.
To date, the SDSS quasars have been
spectroscopically confirmed ``one-at-a time" with large telescopes
(primarily the Apache Point 3.5-m, HET, and Keck).  In the past year
the SDSS multifiber spectrographs have been commissioned
(see Castander~et~al.~2001),
and spectra of more than~2000 quasars have been analyzed at the time
of this writing
(Richards~et~al.~2001, Vanden Berk~et~al.~2001);
this new spectroscopic capability
should lead to a rapid expansion of the
number of known \hbox{$z > 3.5$ quasars.}

\acknowledgments

We would like to thank Ingo Lehmann for his expert help with determining
the X-ray limits for the objects, Pat Hall for useful comments on an early
draft of the manuscript, and the entire HET operations team for
their efforts during the commissioning and early operations of the telescope.
This work was supported in part by National Science Foundation grants
AST99-00703~(DPS and~GTR), AST00-71091~(MAS), and PHY00-70928~(XF).
MAS and XF acknowledge
additional support from the Princeton University
Research Board, and a Porter O.~Jacobus Fellowship.

The Sloan Digital Sky Survey
\footnote{The SDSS Web site \hbox{is {\tt http://www.sdss.org/}.}}
(SDSS) is a joint project of the University of
Chicago, Fermilab, the Institute for Advanced Study, the Japan Participation
Group, the Johns Hopkins University, the Max-Planck-Institute for Astronomy,
New Mexico State University,
Princeton University, the United States Naval Observatory, and the University
of Washington.  Apache Point Observatory, site of the SDSS, is operated by
the Astrophysical Research Consortium.  Funding for the project has been
provided by the Alfred P.~Sloan Foundation, the SDSS member institutions,
the National Aeronautics and Space Administration, the National Science
Foundation, the U.S.~Department of Energy, Monbusho, and the Max Planck
Society.

The Hobby-Eberly Telescope (HET) is a joint project of the University of Texas
at Austin,
the Pennsylvania State University,  Stanford University,
Ludwig-Maximillians-Universit\"at M\"unchen, and Georg-August-Universit\"at
G\"ottingen.  The HET is named in honor of its principal benefactors,
William P. Hobby and Robert E. Eberly.
The Marcario Low Resolution Spectrograph
is named for Mike Marcario of High Lonesome Optics who fabricated 
several optics for the instrument but died before its completion, and is
a joint project of the Hobby-Eberly Telescope partnership and the Instituto de
Astronomia de la Universidad Nacional Autonoma de Mexico.

\clearpage

%
%

%
%
\newpage
\centerline{\bf Figure Captions}

\figcaption{
Finding charts for the quasars; north is up and east to the left, and the
chart is~100$''$ on a side.  All frames are~$i'$ images taken with the SDSS
camera.
\label{fig1}}

\figcaption{
$(a$-$c)$ Spectra of the quasars taken with the Low-Resolution Spectrograph
on the Hobby-Eberly Telescope.  The LRS exposure times and date of observation
are given for each object.
The data have been rebinned \hbox{to 8~\AA\ pixel$^{-1}$.}
The spectral resolution is~20~\AA ;
the unit of flux density is \hbox{AB = 28.0} or
\hbox{$2.29 \times 10^{-31}$ erg cm$^{-2}$ s$^{-1}$ Hz$^{-1}$.}
\label{fig2}}

\figcaption{
Locations of the~113 published SDSS high-redshift quasars in the
\hbox{$(g^*-r^*),(r^*-i^*)$} (left panel)
and \hbox{$(r^*-i^*),(i^*-z^*)$} (right panel)
color-color diagrams.  The quasars are coded by redshift; circles represent
redshifts less than~4.0, triangles are quasars with redshifts between~4.0
and~4.5, and six-pointed stars are quasars with redshifts larger than 4.5.
The solid line is the median track of simulated quasar colors as a function
of redshift from Fan~(1999).
The stellar locus is constructed from stars brighter than $i^*$~=~20.0 in
\hbox{25 square degrees} of SDSS imaging data.
\label{fig3}}

\figcaption{
The redshift distributions of the~113 published SDSS quasars.  Six of the
objects were previously known, but were identified
in SDSS imaging data as high-redshift quasar candidates.
\label{fig4}}
\clearpage

\begin{figure}
\plotfiddle{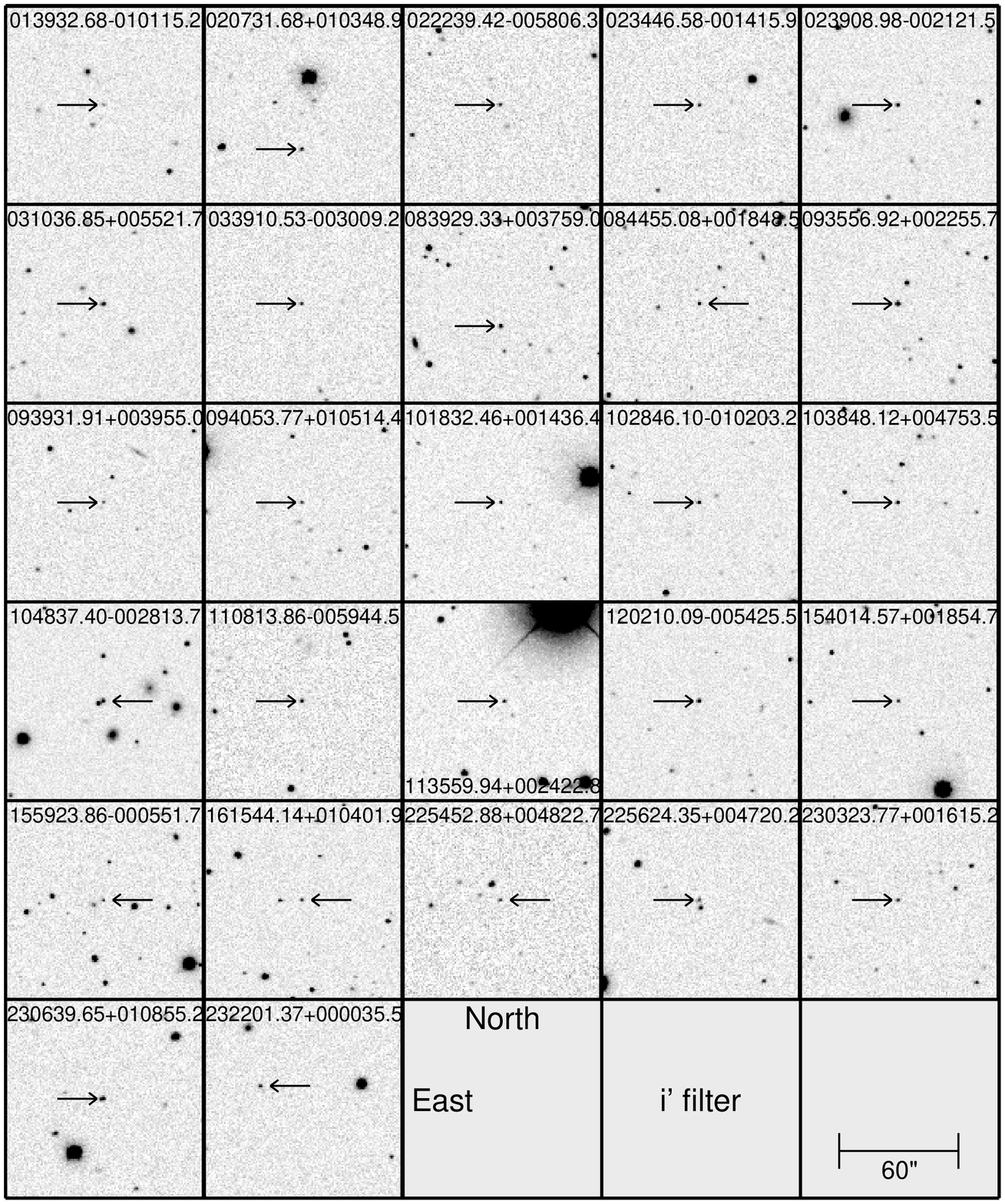}{6.0in}{0.0}{90.0}{90.0}{-270.0}{-150.0}
\end{figure}
\clearpage

\begin{figure}
\plotfiddle{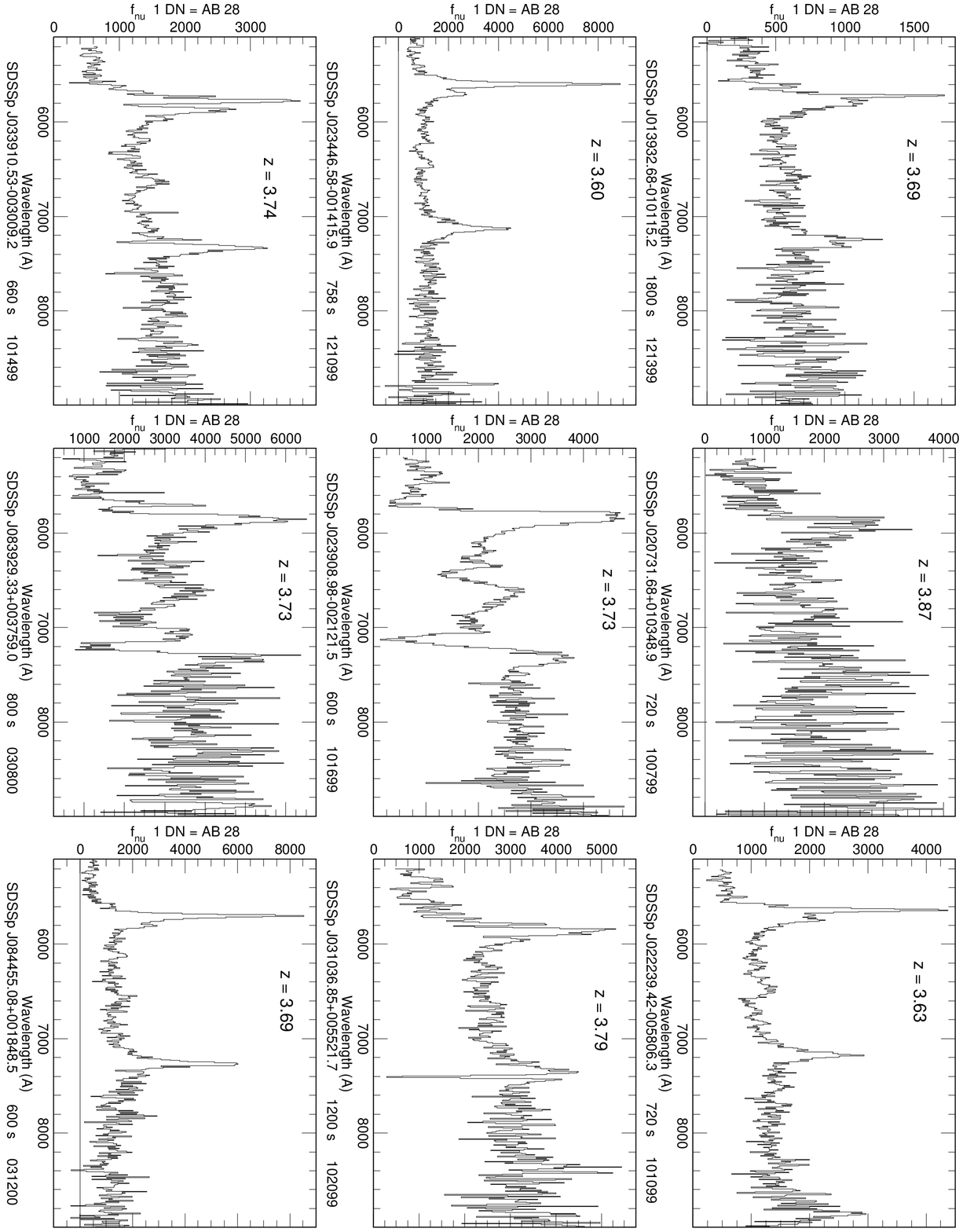}{6.0in}{180.0}{90.0}{90.0}{275.0}{580.0}
\end{figure}
\clearpage

\begin{figure}
\plotfiddle{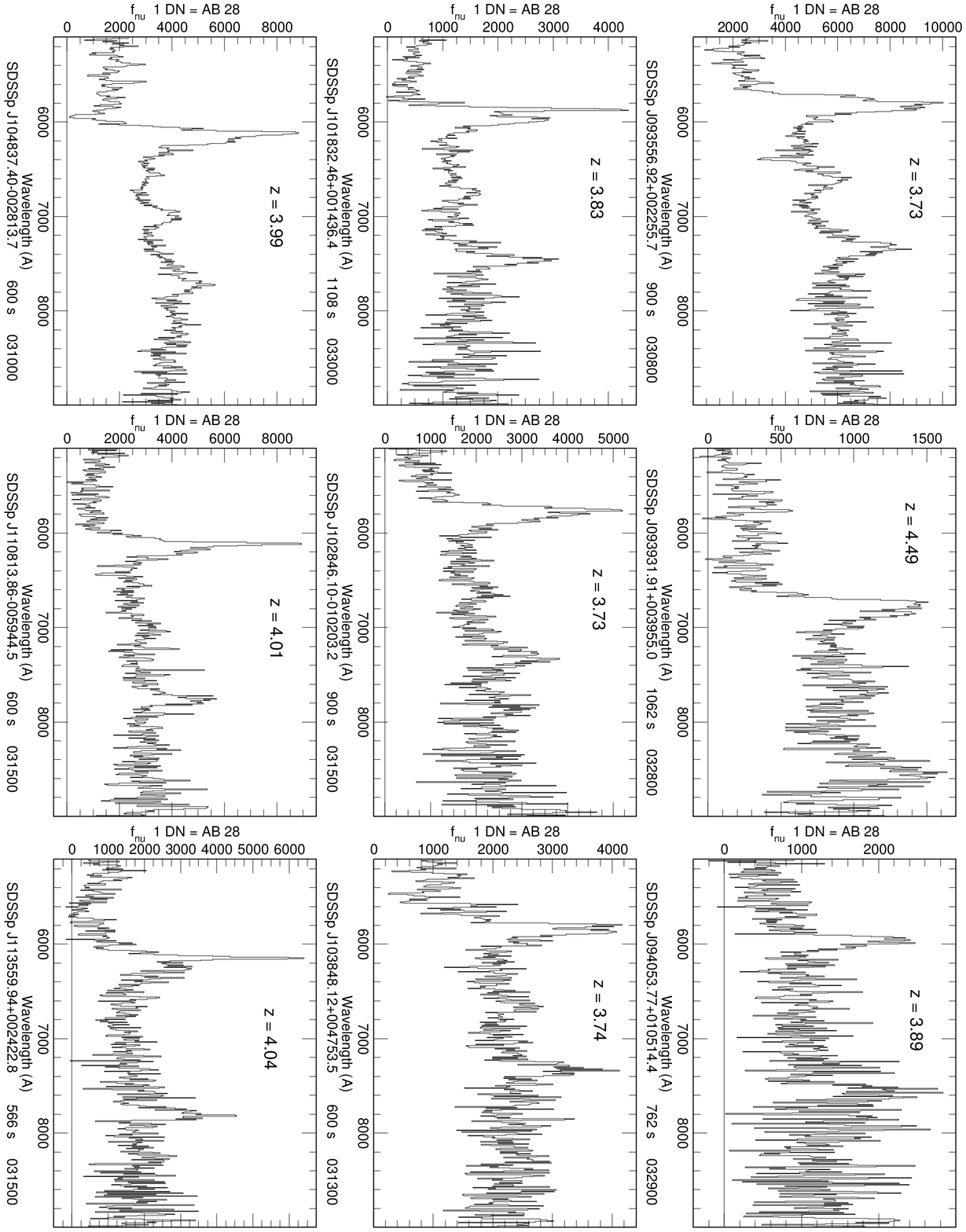}{6.0in}{180.0}{90.0}{90.0}{275.0}{580.0}
\end{figure}
\clearpage

\begin{figure}
\plotfiddle{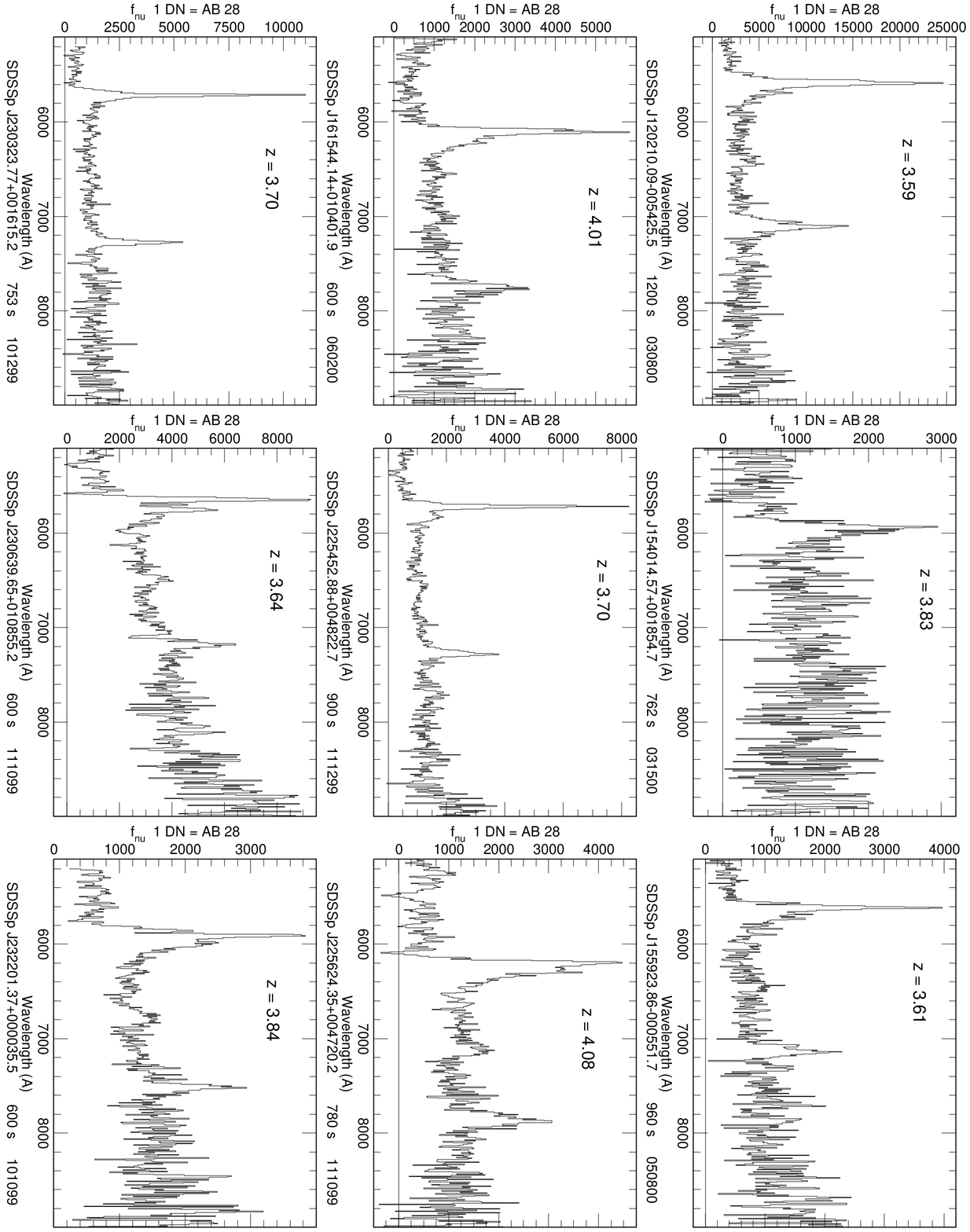}{6.0in}{180.0}{90.0}{90.0}{275.0}{580.0}
\end{figure}
\clearpage

\begin{figure}
\plotfiddle{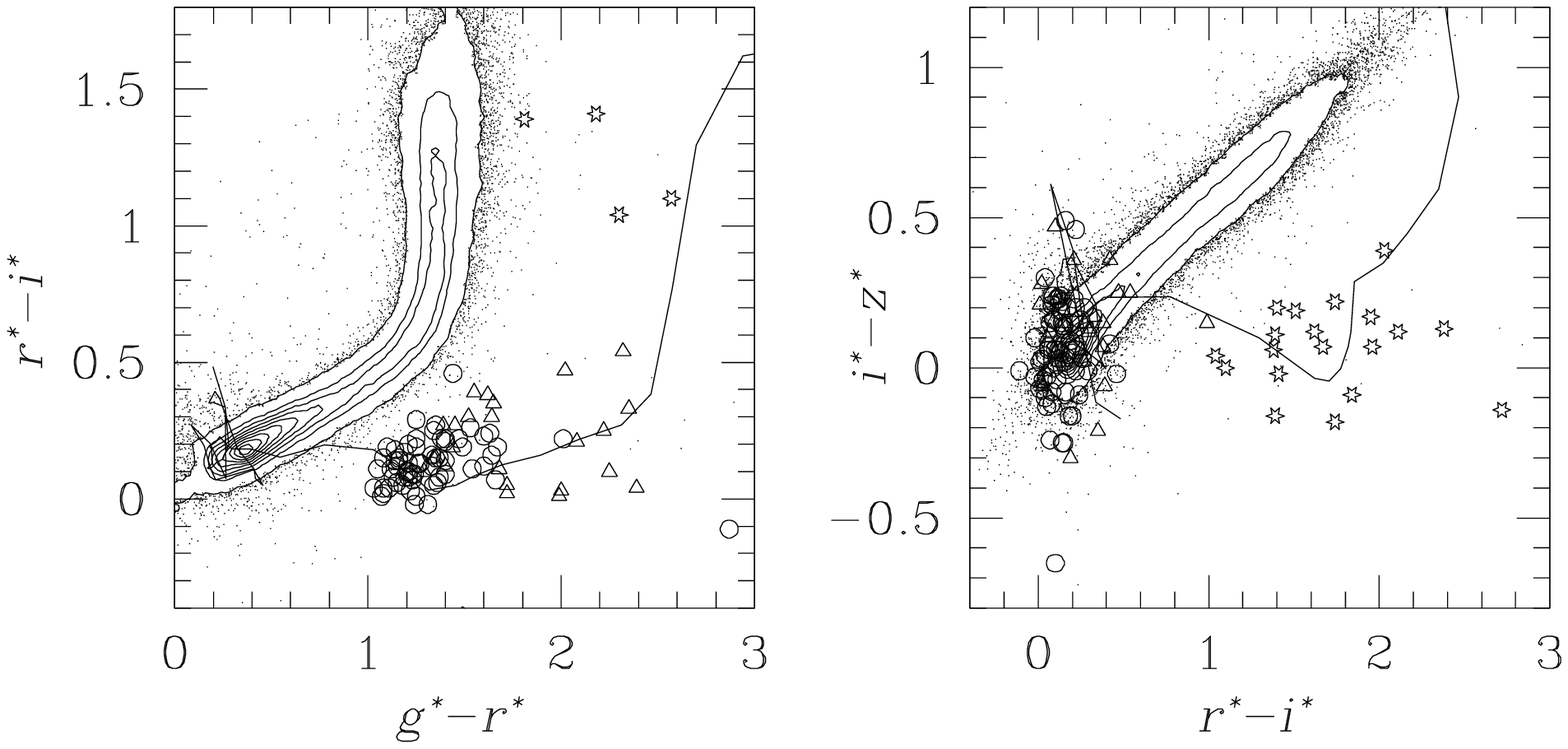}{6.0in}{90.0}{120.0}{120.0}{640.0}{-170.0}
\end{figure}
\clearpage

\begin{figure}
\plotfiddle{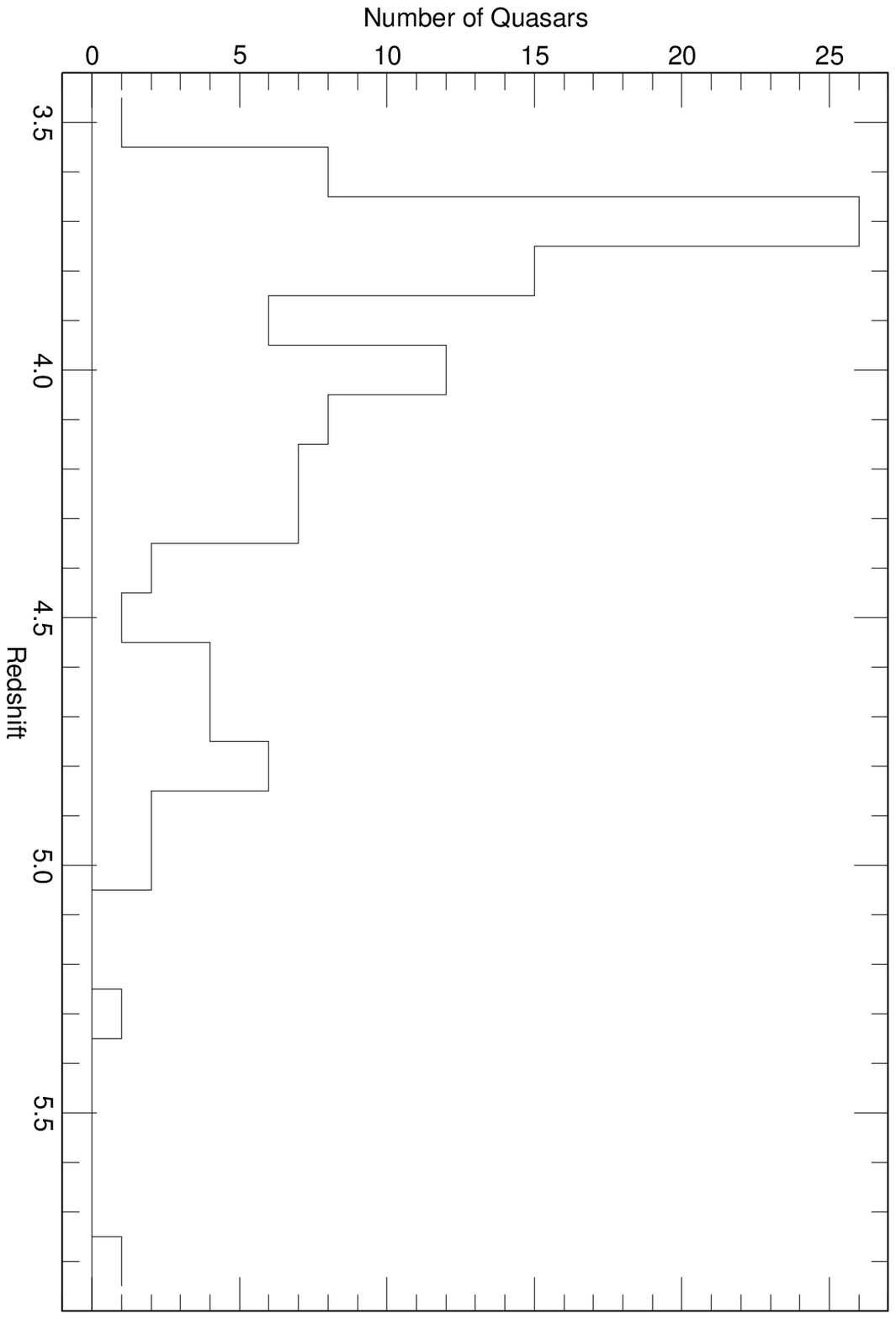}{6.0in}{180.0}{120.0}{120.0}{275.0}{780.0}
\end{figure}
\clearpage

\begin{footnotesize}

\halign{\hskip 12pt
# \hfil \tabskip=1em plus1em minus1em&
\hfil # \hfil &
\hfil # &
\hfil # &
\hfil # &
\hfil # &
\hfil # &
\hfil # \cr
\multispan8{\hfil TABLE 1. Positions and Photometry of SDSS High-Redshift
Quasars \hfil}\cr
\noalign{\bigskip\hrule\smallskip\hrule\medskip}
\hfil Quasar (SDSSp J) \hfil&\hfil Redshift \hfil 
&\hfil $u^*$ \hfil&\hfil $g^*$ \hfil &
\hfil $r^*$ \hfil & \hfil $i^*$ \hfil & \hfil $z^*$ \hfil & \hfil Run \hfil \cr
\noalign{\medskip\hrule\bigskip}
013932.68$-$010115.2  &  3.69 $\pm$ 0.01 & 25.14 $\pm$ 0.85 & 22.09 $\pm$ 0.10
  & 21.01 $\pm$ 0.05 & 20.97 $\pm$ 0.08 & 20.67 $\pm$ 0.23 &  94 \cr
020731.68$+$010348.9  &  3.87 $\pm$ 0.02 & 23.93 $\pm$ 0.41 & 21.20 $\pm$ 0.04
  & 20.04 $\pm$ 0.02 & 19.89 $\pm$ 0.03 & 19.75 $\pm$ 0.11 &  94 \cr
022239.42$-$005806.3  &  3.63 $\pm$ 0.01 & 24.23 $\pm$ 0.59 & 21.31 $\pm$ 0.05
  & 20.23 $\pm$ 0.03 & 20.08 $\pm$ 0.04 & 19.95 $\pm$ 0.14 &  94 \cr
023446.58$-$001415.9  &  3.60 $\pm$ 0.02 & 23.91 $\pm$ 0.64 & 21.40 $\pm$ 0.06
  & 20.24 $\pm$ 0.03 & 20.10 $\pm$ 0.03 & 20.35 $\pm$ 0.15 & 125 \cr
023908.98$-$002121.5* &  3.73 $\pm$ 0.03 & 24.85 $\pm$ 0.78 & 20.99 $\pm$ 0.04
  & 19.62 $\pm$ 0.02 & 19.54 $\pm$ 0.02 & 19.33 $\pm$ 0.07 & 125 \cr
\noalign{\smallskip}
031036.85$+$005521.7  &  3.79 $\pm$ 0.01 & 23.24 $\pm$ 0.54 & 20.99 $\pm$ 0.04
  & 19.50 $\pm$ 0.02 & 19.31 $\pm$ 0.02 & 19.16 $\pm$ 0.06 & 125 \cr
033910.53$-$003009.2  &  3.74 $\pm$ 0.01 & 23.30 $\pm$ 0.57 & 21.30 $\pm$ 0.04
  & 20.09 $\pm$ 0.02 & 19.96 $\pm$ 0.03 & 19.72 $\pm$ 0.10 & 125 \cr
083929.33$+$003759.0* &  3.73 $\pm$ 0.03 & 23.55 $\pm$ 0.49 & 20.52 $\pm$ 0.03
  & 19.28 $\pm$ 0.02 & 19.19 $\pm$ 0.02 & 18.95 $\pm$ 0.04 & 756 \cr
084455.08$+$001848.5  &  3.69 $\pm$ 0.01 & 23.06 $\pm$ 0.36 & 21.34 $\pm$ 0.04
  & 19.94 $\pm$ 0.02 & 19.85 $\pm$ 0.03 & 19.85 $\pm$ 0.09 & 756 \cr
093556.92$+$002255.7  &  3.73 $\pm$ 0.01 & 25.03 $\pm$ 0.44 & 19.83 $\pm$ 0.03
  & 18.69 $\pm$ 0.02 & 18.55 $\pm$ 0.02 & 18.40 $\pm$ 0.04 & 756 \cr
\noalign{\smallskip}
093931.91$+$003955.0  &  4.49 $\pm$ 0.01 & 22.89 $\pm$ 0.37 & 24.32 $\pm$ 0.44
  & 21.61 $\pm$ 0.07 & 20.62 $\pm$ 0.05 & 20.47 $\pm$ 0.14 & 756 \cr
094053.77$+$010514.4  &  3.89 $\pm$ 0.01 & 23.93 $\pm$ 0.43 & 21.70 $\pm$ 0.06
  & 20.45 $\pm$ 0.04 & 20.26 $\pm$ 0.04 & 20.17 $\pm$ 0.13 & 756 \cr
101832.46$+$001436.4  &  3.83 $\pm$ 0.01 & 24.12 $\pm$ 0.51 & 21.63 $\pm$ 0.05
  & 20.23 $\pm$ 0.03 & 20.09 $\pm$ 0.03 & 19.87 $\pm$ 0.09 & 756 \cr
102846.10$-$010203.2  &  3.73 $\pm$ 0.01 & 23.91 $\pm$ 0.50 & 21.02 $\pm$ 0.04
  & 19.69 $\pm$ 0.02 & 19.57 $\pm$ 0.02 & 19.54 $\pm$ 0.07 & 756 \cr
103848.12$+$004753.5  &  3.74 $\pm$ 0.01 & 22.46 $\pm$ 0.19 & 20.77 $\pm$ 0.03
  & 19.64 $\pm$ 0.02 & 19.56 $\pm$ 0.02 & 19.32 $\pm$ 0.05 & 756 \cr
\noalign{\smallskip}
104837.40$-$002813.7  &  3.99 $\pm$ 0.02 & 24.88 $\pm$ 0.57 & 20.87 $\pm$ 0.04
  & 19.27 $\pm$ 0.02 & 19.04 $\pm$ 0.06 & 18.97 $\pm$ 0.13 & 756 \cr
110813.86$-$005944.5  &  4.01 $\pm$ 0.01 & 23.72 $\pm$ 0.44 & 21.01 $\pm$ 0.03
  & 19.56 $\pm$ 0.01 & 19.29 $\pm$ 0.02 & 19.14 $\pm$ 0.05 & 756 \cr
113559.94$+$002422.8  &  4.04 $\pm$ 0.01 & 23.62 $\pm$ 0.35 & 21.37 $\pm$ 0.05
  & 20.03 $\pm$ 0.03 & 19.88 $\pm$ 0.03 & 19.73 $\pm$ 0.08 & 756 \cr
120210.09$-$005425.5  &  3.59 $\pm$ 0.01 & 23.54 $\pm$ 0.58 & 20.10 $\pm$ 0.02
  & 19.05 $\pm$ 0.01 & 18.94 $\pm$ 0.01 & 18.93 $\pm$ 0.04 & 756 \cr
154014.57$+$001854.7  &  3.83 $\pm$ 0.02 & 24.64 $\pm$ 0.52 & 21.75 $\pm$ 0.05
  & 20.42 $\pm$ 0.04 & 20.26 $\pm$ 0.04 & 20.34 $\pm$ 0.17 & 756 \cr
\noalign{\smallskip}
155923.86$-$000551.7  &  3.61 $\pm$ 0.01 & 23.59 $\pm$ 0.46 & 21.79 $\pm$ 0.05
  & 20.65 $\pm$ 0.03 & 20.47 $\pm$ 0.04 & 20.63 $\pm$ 0.18 & 745 \cr
161544.14$+$010401.9  &  4.01 $\pm$ 0.01 & 24.33 $\pm$ 0.62 & 21.74 $\pm$ 0.05
  & 20.34 $\pm$ 0.02 & 20.22 $\pm$ 0.04 & 20.20 $\pm$ 0.13 & 745 \cr
225452.88$+$004822.7  &  3.70 $\pm$ 0.01 & 23.01 $\pm$ 0.55 & 21.52 $\pm$ 0.07
  & 20.26 $\pm$ 0.03 & 20.13 $\pm$ 0.05 & 19.94 $\pm$ 0.16 &  94 \cr
225624.35$+$004720.2  &  4.08 $\pm$ 0.01 & 23.75 $\pm$ 0.80 & 21.96 $\pm$ 0.10
  & 20.28 $\pm$ 0.04 & 20.17 $\pm$ 0.07 & 19.94 $\pm$ 0.16 &  94 \cr
230323.77$+$001615.2  &  3.70 $\pm$ 0.01 & 25.31 $\pm$ 0.39 & 21.22 $\pm$ 0.05
  & 20.11 $\pm$ 0.03 & 20.07 $\pm$ 0.04 & 20.03 $\pm$ 0.20 &  94 \cr
\noalign{\smallskip}
230639.65$+$010855.2  &  3.64 $\pm$ 0.01 & 23.91 $\pm$ 0.51 & 20.59 $\pm$ 0.03
  & 19.25 $\pm$ 0.01 & 19.00 $\pm$ 0.02 & 18.84 $\pm$ 0.06 &  94 \cr
232201.37$+$000035.5  &  3.84 $\pm$ 0.01 & 24.55 $\pm$ 0.60 & 21.34 $\pm$ 0.05
  & 20.19 $\pm$ 0.03 & 20.07 $\pm$ 0.04 & 20.01 $\pm$ 0.15 &  94 \cr
\noalign{\medskip\hrule}}
\medskip\noindent
Notes: Positions are in J2000.0 coordinates.
An asterisk following an object name indicates a BAL quasar.
Photometry is reported in terms of asinh magnitudes; see 
Lupton, Gunn, \& Szalay~(1999) for details.  In this system, zero flux
corresponds to 23.24, 24.91, 24.53, 23.89, and~22.47 in $u^*$, $g^*$,
$r^*$, $i^*$, and~$z^*$, respectively.

\end{footnotesize}
\clearpage
\halign{\hskip 12pt
# \hfil \tabskip=1em plus1em minus1em&
\hfil # \hfil &
\hfil # \hfil &
\hfil # \hfil &
\hfil $#$ \hfil &
\hfil $#$ &
\hfil $#$ \hfil \cr
\multispan6{\hfil TABLE 2. Properties of SDSS High-Redshift Quasars
\hfil}\cr
\noalign{\bigskip\hrule\smallskip\hrule\medskip}
\hfil Quasar (SDSSp J) \hfil&\hfil Redshift \hfil & $E(B-V)$
&\hfil $AB_{1450}$ \hfil & \hfil \alpha & \hfil \Delta L\alpha
& \hfil M_{1450} \hfil \cr
\noalign{\medskip\hrule\bigskip}
013932.68$-$010115.2  &  3.69 & 0.034 & 21.08 & -0.93 &  5.1& -24.65 \cr
020731.68$+$010348.9  &  3.87 & 0.026 & 20.00 & -0.96 & -2.3& -25.80 \cr
022239.42$-$005806.3  &  3.63 & 0.036 & 20.27 & -0.91 &  1.2& -25.43 \cr
023446.58$-$001415.9  &  3.60 & 0.022 & 20.40 & -0.89 &  1.0& -25.29 \cr
023908.98$-$002121.5* &  3.73 & 0.028 & 19.53 & -0.92 &  8.4& -26.21 \cr
\noalign{\smallskip}
031036.85$+$005521.7  &  3.79 & 0.115 & 19.18 & -0.76 &  8.1& -26.59 \cr
033910.53$-$003009.2  &  3.74 & 0.104 & 19.90 & -1.14 &  2.6& -25.85 \cr
083929.33$+$003759.0* &  3.73 & 0.042 & 19.14 & -0.84 & 22.1& -26.60 \cr
084455.08$+$001848.5  &  3.69 & 0.035 & 20.05 & -1.11 & -0.6& -25.68 \cr
093556.92$+$002255.7  &  3.73 & 0.050 & 18.61 & -1.03 & 15.2& -27.13 \cr
\noalign{\smallskip}
093931.91$+$003955.0  &  4.49 & 0.065 & 20.52 & -0.42 & 13.4& -25.51 \cr
094053.77$+$010514.4  &  3.89 & 0.097 & 20.29 & -0.28 & -1.8& -25.52 \cr
101832.46$+$001436.4  &  3.83 & 0.043 & 20.19 & -1.09 &  0.5& -25.59 \cr
102846.10$-$010203.2  &  3.72 & 0.052 & 19.72 & -1.17 &  1.3& -26.02 \cr
103848.12$+$004753.5  &  3.74 & 0.060 & 19.54 & -0.65 & 17.5& -26.21 \cr
\noalign{\smallskip}
104837.40$-$002813.7  &  3.99 & 0.040 & 19.08 & -1.19 & 10.3& -26.77 \cr
110813.86$-$005944.5  &  4.01 & 0.047 & 19.35 & -1.01 &  4.0& -26.51 \cr
113559.94$+$002422.8  &  4.04 & 0.019 & 19.96 & -0.95 &  4.9& -25.91 \cr
120210.09$-$005425.5  &  3.59 & 0.025 & 19.25 & -0.91 &  2.0& -26.43 \cr
154014.57$+$001854.7  &  3.83 & 0.093 & 20.20 & -0.57 & 13.7& -25.59 \cr
\noalign{\smallskip}
155923.86$-$000551.7  &  3.61 & 0.126 & 20.51 & -1.58 &  2.8& -25.18 \cr
161544.14$+$010401.9  &  4.01 & 0.091 & 20.18 & -0.88 &  1.0& -25.68 \cr
225452.88$+$004822.7  &  3.70 & 0.087 & 20.22 & -1.11 &  1.5& -25.51 \cr
225624.35$+$004720.2  &  4.08 & 0.059 & 20.14 & -0.25 &  4.4& -25.75 \cr
230323.77$+$001615.2  &  3.70 & 0.049 & 20.18 & -0.93 &  1.4& -25.55 \cr
\noalign{\smallskip}
230639.65$+$010855.2  &  3.64 & 0.053 & 19.15 & -1.58 &  1.3& -26.56 \cr
232201.37$+$000035.5  &  3.84 & 0.043 & 20.13 & -1.14 &  4.5& -25.66 \cr
\noalign{\medskip\hrule}}

\medskip\noindent
Notes: An asterisk following an object name indicates a BAL quasar.
Galactic reddening values from Schlegel, Finkbeiner, \& Davis (1998), and
a standard reddening law $R$~=~3.1 were used to calculate~$AB_{1450}$.  The
absolute magnitude $M_{1450}$ was calculated assuming
\hbox{$H_0$ = 50 km s$^{-1}$ Mpc$^{-1}$,}
$\Omega_0$~=~1, and~$\Lambda$~=~0.
Absolute~$B$ magnitudes can be
calculated from \hbox{$M_B = M_{1450} + 1.21 \alpha + 0.12$,} where~$\alpha$
is the spectral power law index \hbox{($f_{\nu} \propto \nu^{\alpha}$)}
between the rest frame ultraviolet and blue;
\hbox{$M_B = M_{1450} - 0.49$} for \hbox{$\alpha = -0.5$.}

\end{document}